\documentstyle[11pt]{article}

\catcode`\@=11
\long\def\@makefntext#1{
\protect\noindent \hbox to 3.2pt {\hskip-.9pt
$^{{\ninerm\@thefnmark}}$\hfil}#1\hfill}                %CAN BE USED

 \def\@makefnmark{\hbox to 0pt{$^{\@thefnmark}$\hss}}  %ORIGINAL

\def\ps@myheadings{\let\@mkboth\@gobbletwo
\def\@oddhead{\hbox{}
\rightmark\hfil\ninerm\thepage}
\def\@oddfoot{}\def\@evenhead{\ninerm\thepage\hfil
\leftmark\hbox{}}\def\@evenfoot{}
\def\sectionmark##1{}\def\subsectionmark##1{}}

\newcounter{sectionc}\newcounter{subsectionc}\newcounter{subsubsectionc}
\renewcommand{\section}[1] {\vspace{0.6cm}\addtocounter{sectionc}{1}
\setcounter{subsectionc}{0}\setcounter{subsubsectionc}{0}\noindent
    {\bf\thesectionc. #1}\par\vspace{0.4cm}}
\renewcommand{\subsection}[1] {\vspace{0.6cm}\addtocounter{subsectionc}{1}
    \setcounter{subsubsectionc}{0}\noindent
    {\it\thesectionc.\thesubsectionc. #1}\par\vspace{0.4cm}}
\renewcommand{\subsubsection}[1] {\vspace{0.6cm}\addtocounter{subsubsectionc}{1}
    \noindent {\rm\thesectionc.\thesubsectionc.\thesubsubsectionc.
    #1}\par\vspace{0.4cm}}

\newcounter{appendixc}
\newcounter{subappendixc}[appendixc]
\newcounter{subsubappendixc}[subappendixc]

\renewcommand{\appendix}[1] {\vspace{0.6cm}
    \refstepcounter{appendixc}
    \setcounter{figure}{0}
    \setcounter{table}{0}
    \setcounter{equation}{0}
    \renewcommand{\thefigure}{\Alph{appendixc}.\arabic{figure}}
    \renewcommand{\thetable}{\Alph{appendixc}.\arabic{table}}
    \renewcommand{\theappendixc}{\Alph{appendixc}}
    \renewcommand{\theequation}{\Alph{appendixc}.\arabic{equation}}
    \noindent{\bf Appendix \theappendixc #1}\par\vspace{0.4cm}}

\renewenvironment{thebibliography}[1]
    {\begin{list}{\arabic{enumi}.}
    {\usecounter{enumi}\setlength{\parsep}{0pt}
\setlength{\leftmargin 1.25cm}{\rightmargin 0pt}
     \setlength{\itemsep}{0pt} \settowidth
    {\labelwidth}{#1.}\sloppy}}{\end{list}}

\newcounter{itemlistc}
\newcounter{romanlistc}
\newcounter{alphlistc}
\newcounter{arabiclistc}

\newcommand{\fcaption}[1]{
    \refstepcounter{figure}
    \setbox\@tempboxa = \hbox{\tenrm Fig.~\thefigure. #1}
    \ifdim \wd\@tempboxa > 6in
       {\begin{center}
    \parbox{6in}{\tenrm\baselineskip=12pt Fig.~\thefigure. #1}
        \end{center}}
    \else
         {\begin{center}
         {\tenrm Fig.~\thefigure. #1}
          \end{center}}
    \fi}

\newcommand{\tcaption}[1]{
    \refstepcounter{table}
    \setbox\@tempboxa = \hbox{\tenrm Table~\thetable. #1}
    \ifdim \wd\@tempboxa > 6in
       {\begin{center}
    \parbox{6in}{\tenrm\baselineskip=12pt Table~\thetable. #1}
        \end{center}}
    \else
         {\begin{center}
         {\tenrm Table~\thetable. #1}
          \end{center}}
    \fi}

\def\@citex[#1]#2{\if@filesw\immediate\write\@auxout
    {\string\citation{#2}}\fi
\def\@citea{}\@cite{\@for\@citeb:=#2\do
    {\@citea\def\@citea{,}\@ifundefined
    {b@\@citeb}{{\bf ?}\@warning
    {Citation `\@citeb' on page \thepage \space undefined}}
    {\csname b@\@citeb\endcsname}}}{#1}}

\newif\if@cghi
\def\cite{\@cghitrue\@ifnextchar [{\@tempswatrue
    \@citex}{\@tempswafalse\@citex[]}}
\def\citelow{\@cghifalse\@ifnextchar [{\@tempswatrue
    \@citex}{\@tempswafalse\@citex[]}}
\def\@cite#1#2{{$\null^{#1}$\if@tempswa\typeout
    {IJCGA warning: optional citation argument
    ignored: `#2'} \fi}}

\def\fnt#1#2{\footnotetext{\kern-.3em
    {$^{\mbox{\sevenrm #1}}$}{#2}}}

 1
 1
 1
\font\tenrm=cmr10

\font\ninerm=cmr9

\begin{document}
%\draft
%\title{{\bf  Vacuum decay via lorentzian wormholes}}
%\author{{\bf J.L. ROSALES }\\
%       {\em Fakult\"at f\"ur Physik,\\
%         Universit\"at Freiburg, \\
%         Hermann-Herder St. 3, D-79104 Freiburg im Breisgau, Germany.}}
%\date{\today}
%\begin{center}
%\maketitle
%\begin{abstract}

%\end{abstract}

\vspace{2 cm}
\begin{flushright}
FR-THEP-NR 97/7 \\ gr-qc/9703017
\end{flushright}

\vspace{15 mm}
\begin{center}
\large{{\bf VACUUM DECAY VIA LORENTZIAN WORMHOLES}}
%\end{center}

\vspace{10 mm}
\hspace{2 cm}  J.L. ROSALES \footnote{E-mail: rosales@phyq1.physik.uni-freiburg.de}
\footnote{To my mother}

\hspace{2 cm}
     {\em Fakult\"at f\"ur Physik, Universit\"at Freiburg,}

\hspace{2 cm} {\em Hermann-Herder-Strasse 3, D-79104 Freiburg, Germany}

\vspace{15 mm}

\end{center}

\begin{quote}
\begin{center}
                Abstract
\end{center}
We speculate about the spacetime description
due to the presence of Lorentzian wormholes
(handles in spacetime joining two distant regions
or other universes)
in quantum gravity.
The semiclassical rate of production of these Lorentzian
wormholes in  Reissner - Nordstr\"om spacetimes is calculated
as a result of the spontaneous decay of vacuum due to a
real tunneling configuration.
In the magnetic case it only depends on the
value of the field theoretical fine structure constant.
We predict that the quantum probability corresponding to the nucleation of
such geodesically complete spacetimes should be actually
negligible in our physical Universe.

\end{quote}
\vspace{3 mm}

\section{Introduction}

\noindent This paper is devoted to the real tunneling configurations and
spontaneous nucleation  of vacuum into Lorentzian wormholes as
a possibility to study the quantum gravity foam according
to the seminal considerations of Wheeler \cite{kn:Wheeler} and
Hawking\cite{kn:Hawking1}.
We will present an  extension of the quantum mechanical
analysis of a minisuperspace that describes wormholes in
the Reissner-Nordstr\"om background spacetime;
this model has been already studied in details by Visser \cite{kn:Visser1}
\cite{kn:Visser2}. Wheeler-DeWitt  equation corresponding to
the quantisation of the wormhole dynamic variable, the throat
radius, exibits, in this model, the form of an alternative
finite differences relationship.
Here we will develop the semiclassical
approximation in order to construct from that relationship a
Wheeler-DeWitt equation
which is equivalent to the first one  developed by Visser
up to the smallest order in the WKB approximation.
The solutions are
quantum instantons and their amplitude of "probability" is dominated
by the corresponding Euclidean signature action $ P \sim \exp(-2I)$.
As a result of this, we will develop an interpretation of the
solutions in terms of the tunneling rate corresponding to
the nucleation "from nothing"  of a lorentzian wormhole
of a given charge and mass\cite{kn:Berezinal}.
The calculation procedure is enterely equivalent to that from
minisuperspace models in quantum cosmology.
There, the creation of the Universe "ex nihilo" is due
to the presence of an effective cosmological constant and the result may also
be considered as a sort of classical change of signature or a bounce
in  spacetime (i.e., the transition between two solutions with the same boundary
conditions having different actions).
In quantum cosmology, a tunneling solution of the
Wheeler-DeWitt equation in minisuperspace with a possitive
cosmological constant would  represent
the quantum rate of production corresponding to a spacetime
of the type of a round Euclidean
sphere joined on an equator to the Lorentzian de Sitter space at
its radius of maximum contraction \cite{kn:Gibbons1}.

The spontaneous decay will be regulated by an
arbitrarily small negative mass paramenter.
This is required  for any solution of general relativity
which undertake a change of topology \cite{kn:Tipler} (actually, Tipler's theorem
is the negative of the previous sentence, i.e., given the possitiveness
of the energy-momentum tensor, there exits no classical change in
the topology of spacetime). A change of topology could only exist as a quantum
phenomenon but still, since we require Einstein's equations in order
to formulate the ground theory,
we would need  the violation of the energy condition,
at least minimally.
The important issue is to notice that we obtain results according
to the semiclassical theory but that they would, very likely, be of value
in order to investigate more profound properties of the expected quantum theory of
gravity.

In section 2, we review the dynamic of  Lorentzian wormholes
connecting two asymptotically flat Reissner-Nordstr\"om spacetimes
after the approach of Visser. In section 3, we derive, following semiclassical
considerations, an alternative Wheeler-DeWitt equation whose
solutions exhibit a "tunneling from nothing" configuration
allowing for a  simpler interpretation of  Visser equations in
the semiclassical limit.
We also  proof, in section 3, the consistence of the quasiclassical
expressions  and the qualitative behaviour of the solutions.
The vacuum decay rate via Lorentzian
wormholes is obtained from the tunneling configuaration;
in the magnetic case, the rate can be written as an expression
in terms of  the field theoretical fine structure constant.
We summarize our conclusions in section 4.

\section{Lorentzian wormholes}

Wormholes are handles in the spacetime topology
linking widely separated regions of the Universe, or bridges joining two
different spacetimes.
Thus, a Lorentzian wormhole consists on the union of two copies of identical
asymptotically flat four dimensional regions $\Omega_{1}$ and $\Omega_{2}$
representing those regions.
Technically, one is left with two geodesically incomplete manifolds with boundaries
given by the  hypersurfaces $\partial \Omega_{1}$ and $\partial \Omega_{2}$.
Now, identify these two hypersurfaces. The resulting spacetime $M$ is
geodesically complete and posseses two asymptotically flat regions
connected by a wormhole. The throat of the wormhole is at $\partial \Omega$.
At the throat the Riemann curvature tensor is proportional to
a delta function while it vanishes exactly outside in $\Omega_{1}$ and
$\Omega_{2}$. Since we are looking for solutions of the Einstein equations
with that topology, we require the "junction conditions" which were already
developed by Israel \cite{kn:Israel}.
The required junction conditions are most conveniently derived by
introducing gaussian normal coordinates
in the neighbourhood of the hypersurface $\partial \Omega$:
\begin{equation}
ds^2=\sigma d\eta^2+g_{ij}(\eta,x^{i})dx^{i}dx^{j},
\end{equation}
where $\sigma=1, (-1)$ if $\partial \Omega$ is spacelike (timelike).
The gaussian coordinate $\eta$ parametrizes the proper distance measured
perpendicularly through $\partial \Omega=\{(\eta, x^{i}) | \eta=0\}$. Define
the surface stress-tensor on $\partial \Omega$ by
\begin{equation}
S^{\mu}_{\nu}=\lim_{\varepsilon \rightarrow 0}\int_{-\varepsilon}^{\varepsilon} d\eta
T^{\mu}_{\nu},
\end{equation}
where $T^{\mu}_{\nu}$ contains all possible sources. We wish the geometry
($g_{ij}$) to be continuous at the boundary, but the metric need not be
differentiable there. Then the metric junction conditions, derived by
integrating the Einstein equations across $\partial \Omega$ are
\begin{equation}
S^{\eta}_{\eta}=0,
\end{equation}
\begin{equation}
S^{\eta}_{i}=0,
\end{equation}
and,
\begin{equation}
\sigma[\Delta K^{i}_{j}-\delta Tr(\Delta K)]=-8\pi S^{i}_{j},
\end{equation}
where
\begin{equation}
\Delta K^{i}_{j}=\lim_{\varepsilon \rightarrow 0}(K^{+i}_{j}-K^{-i}_{j})
\end{equation}
is the "jump" in the extrinsic curvature of the layer in going from the
$-\varepsilon$ to the $+\varepsilon$ side.
The Ricci tensor at the junction
can be calculated in terms of the extrinsic curvature
\cite{kn:Visser1}, \cite{kn:Visser3}
\begin{equation}
K^{i}_{j}=\frac{1}{2}g^{ik}\frac{\partial g_{kj}}{\partial \eta}|_{\eta=0}\mbox{.}
\end{equation}
Here $\eta$ also denotes the normal coordinate to the throat, while $\tau$ will
hereafter denote
the proper time along the throat, i.e., the geometry of the boundary is
given by the metric
\begin{equation}
ds^2=-d\tau^2+a(\tau)^{2}d\Omega_{3}^2,
\end{equation}
where $a(\tau)$ is the throat radius and $d\Omega_{3}^2$ states for the metric
of the unit three - sphere.

Let us consider the Ricci tensor
from the Reissner-Nordstr\"om geometry almost everywhere but on the throat:

\begin{equation}
R^{\mu}_{\nu}=R^{(RN)\mu}_{\nu}-2\left( \begin{array}{c} K^{i}_{j}(x) \,\,\,\, 0 \\ 0 \,\,\,\, K(x) \end{array} \right)\delta(\eta)
\mbox{;}\,\,\,\,\,\,
\end{equation}
So that the Einstein-Hilbert action reduces to
\begin{equation}
S_{G}=\frac{1}{16\pi}\int_{M}(-g_{4})^{1/2}R
=-\frac{1}{4\pi}\int_{\partial \Omega}(g_{3})^{1/2} K \mbox{.}
\end{equation}

By considering  the joining conditions it is possible to write the
extrinsic curvature in terms of the parameters $a$ and $\dot{a}=d a/d\tau$
(the  proper velocity of the throat).
Upon taking into account the electromagnetic action
within the regions $\Omega_{1}$ and $\Omega_{2}$ one finally
obtains for the effective action of the throat
\begin{equation}
S_{eff}=2\int\{a\dot{a} \sinh^{-1}(a\dot{a}/h(a)^{1/2})-(h(a)+a^2\dot{a}^2)^{1/2}\}d\tau \mbox{,}
\end{equation}
where,
\begin{equation}
h(a)\equiv a^2-2ma+ q^2,
\end{equation}
$q$ and $m$ being respectively the charge and the mass of the corresponding
Reissner-Nordstr\"om metrics.

As stated above, let us now assume the presence of a
certain amount of mass $w$ on the throat, the mass of the dust shell. This is a
constant of motion. The matter Lagrangian reduces to $L_{m}=-w$, and the total
lagrangian would be
\begin{equation}
L=2\{a\dot{a} \sinh^{-1}(a\dot{a}/h(a)^{1/2})-(h(a)+a^2\dot{a}^2)^{1/2}\}-w \mbox{.}
\end{equation}

The Hamiltonian is given by
\begin{equation}
H(p,a)=2 h(a)^{1/2}\cosh[\frac{p}{2a}]+w=2(h(a)+a^2\dot{a}^2)^{1/2}+w \mbox{,}
\end{equation}
and the constraint $H=0$ leads to the equation of motion for the throat radius
\begin{equation}
h(a)+a^2\dot{a}^2 -\frac{w^2}{4}=0 \mbox{.}
\end{equation}

The Wheeler-DeWitt equation for the wave function of the dynamic variable $a$
in this minisuperspace  is written as follows
\begin{equation}
\{2 h(a)^{1/2-s}\cos[\frac{1}{2aM^2}\frac{\partial}{\partial a} ] h(a)^{s}+w \}\psi(a)=0 \mbox{,}
\end{equation}
where $s$ is a constant factor representing part of the factor ordering
ambiguity and $M^2$ denotes Planck's mass squared.
On the other hand,  Visser
\cite{kn:Visser2} takes $s=1/4$ and restricts the solutions to the
space of integrable functions in $[0,\infty)$. Here, let us select simply
$s=0$. In this case,
upon substituting the variable $a^2=x$, we have,
instead, a finite differences equation \cite{kn:Hochberg}, namely
\begin{equation}
\psi(x+i/M^2)+\psi(x-i/M^2)+\frac{w}{h(x)^{1/2}}\psi(x)=0 \mbox{;}
\end{equation}
which has the remarkable property that  it is complex conjugation invariant if
and only if $h(a)^{1/2}$ is real,i.e.,  in case that $m=q$. In other
words, the reality of the wave function implies that the asymptotically
flat spacetime is just the extreme Reissner-Nordstr\"om  black hole.
Since time is absent from this quantum theory , such a complex  conjugation
invariance could be required by the consistence of the quantisation procedure
(although, quantum gravity might not satisfy this requirement strictly
\cite{kn:Rosales}). On the other hand, the existence of a
finite differences equation seems to be a fact related with the speculations
of Wheeler\cite{kn:Wheeler} in the sense that spacetime might not be considered as
classical on scales of the order of the Planck length ($a\sim M^{-1}$), i.e.,
in our terms,
that the continuous character of spacetime can only be
recovered, we will see, as an approximated phenomenon having
a semiclassical meaning.

\newpage

\section{Semiclassical theory}

Let us now obtain the semiclassical approximation to  Eq. (17). Our hope
will be, therefore, to recover a suitable continuous limit in
case that $M a\gg 1$.
Notice that Eq. (17) may also be written as follows
\begin{equation}
2xM^2\{\psi(x+i/M^2)+\psi(x-i/M^2)-2\psi(x)\} +\frac{M^2 V^{0}(x)}{2}\psi(x)=0 \mbox{,}
\end{equation}
where
\begin{equation}
V^{0}(x)=4(2x+\frac{w x}{h(x)^{1/2}})=4(2a^2+\frac{wa^2}{h(a)^{1/2}}) \mbox{,}
\end{equation}
and that, for $M^{-2}/x\ll 1$ (radius of the throat much larger than the Planck
length), we have as a first approximation \cite{kn:Berezin}
\begin{equation}
2xM^2\{\psi(x+i/M^2)+\psi(x-i/M^2)-2\psi(x)\}\approx
-\frac{a}{2M^2}\frac{\partial}{\partial a}a^{-1}\frac{\partial }{\partial a}\psi(a) \mbox{.}
\end{equation}
Therefore, the wave function in Eq. (18) is  approximated by the solution
of
\begin{equation}
-\frac{a}{2M^2}\frac{\partial}{\partial a}a^{-1}\frac{\partial }{\partial a}\hat{\psi}(a)+
\frac{M^2 V^{0}(a)}{2}\hat{\psi}(a)=0 \mbox{.}
\end{equation}

For $w=-|w|=-\xi<0$, there exits a potential barrier between $a=0$ and
$a=m-\xi/2$ representing the  nucleation of a wormhole "from nothing"
in a way  anagously to the cosmological
case. This would change the causal structure of spacetime and, therefore,
would also require  the violation of the energy conditions according
to Tipler's theorem \cite{kn:Tipler}. In  case that $w>0$, there would exits no
nucleation and there should be no causal violations in spacetime
(the  wave function is strongly peaked about $a=0$).

On the other hand, we may obtain a more rigurous approximation for the
solutions of Eq. (18). To see this, let us write
\begin{equation}
\psi(a)=\exp[M^{2} S(a)] \mbox{,}
\end{equation}
then, the finite differences equation is approximated up to order $O(M^{-4})$
by
\begin{equation}
-\frac{a^p}{2M^2}\frac{\partial}{\partial a}a^{-p}\frac{\partial }{\partial a}\psi(a)+
\frac{M^2V(a)}{2}\psi(a)=0 \mbox{,}
\end{equation}
with $p=1$ and $V(a)$ given by the series
\begin{equation}
V(a)=V^{0}(a)+8a^2\sum_{k=4}^{\infty}\frac{(-1)^{k}}{(2a)^{k}k!}
(\frac{\partial S}{\partial a})^{k} \mbox{.}
\end{equation}

Now, in the semiclassical approximation,
\begin{equation}
\frac{\partial S}{\partial a}= V(a)^{1/2} +O(1/(M^2)) \mbox{,}
\end{equation}
and, after replacing Eq. (25) in Eq. (24) and suming the series we get
\begin{equation}
V(a)=V^{0}(a)+8a^2[\cos(\frac{V(a)^{1/2}}{2a})-1+ \frac{1}{2}(\frac{V(a)^{1/2}}{2a})^2 ] \mbox{,}
\end{equation}
or {\footnote {\it Notice that, by using Eq. (15), $i\partial L/\partial \dot{a}=V(a)^{1/2}$,
and, upon doing a Wick rotation, these are just
the semiclassical instantons to which our solutions in Eq. (23)
should be correlated about. This is consistent with the  approximations.}}
\begin{equation}
V(a)^{1/2}=2 a \cos^{-1}(\frac{\xi}{2h^{1/2}})) \mbox{.}
\end{equation}
This is equivalent to taken
\begin{equation}
V(a)^{1/2}=a\pi[1-\frac{2}{\pi}\tan^{-1}(\frac{\xi}{(4h-\xi^2)^{1/2}})] \mbox{,}
\end{equation}
if $h(a)=(m-a)^2$, the existence of a barrier
between  $a=0$ ("no wormhole") and $a=m-\xi/2$ is now evident.

In what follows, we will take simply $M=1$,
using the semiclassical theory just in the sense that we
expect $m\gg \xi\sim 1$.

On the other hand,
Eq. (23) will exhibit the same qualitative behaviour than
\begin{equation}
-\frac{a}{2}\frac{\partial }{\partial a} a^{-1}
\frac{\partial }{\partial a}\psi(a)+ \frac{1}{2}a^2
\pi^2[1-\frac{\xi}{2(m-a)}]\psi(a)=0 \mbox{;}
\end{equation}
it has a pole at $a=m^-$. In order to get consistence with the semiclassical
approximations we should investigate the behaviour
of the wave function in that limit; moreover, if we were interested
on computing the tunneling rate between $a=0$ and $a=m^-$
from the known quasiclassical expressions,
there should exist a suitable set of coordinates
which contains the barrier and an asymptotically free region that
map  the pole to infinity within such an asymptotic region.
In this case we require
$\lim_{a\rightarrow m^-}\psi(a)=0$.

In order to deal with the singularity let us use the variable
$\zeta=\frac{1}{m-a}$.
We are interested on the qualitative behaviour of the solutions of
Eq. (29) for $m\zeta \gg 1$.

Define
\begin{equation}
\psi\equiv \phi \frac{(m\zeta-1)^{1/2}}{\zeta^{3/2}} \mbox{,}
\end{equation}
so that,
\begin{equation}
\phi^{''}(\zeta) -\tilde{V}(\zeta)\phi(\zeta)=0      \mbox{;}
\end{equation}
where,
\begin{equation}
\tilde{V}(\zeta)=\frac{3}{4\zeta^{2}(m\zeta-1)^{2}}
-\frac{\pi^{2}(\xi\zeta-2)(m\zeta-1)^{2}}{2\zeta^6}         \mbox{.}
\end{equation}
Far from the turning point at $\zeta\sim 2/\xi$
we may write the approximate asymptotic regime for $\phi(\zeta)$,
\begin{eqnarray*}
\phi_{m\zeta\sim 1}\sim (m\zeta-1)^{-1/2} \mbox{,}
\end{eqnarray*}
\begin{equation}
\phi_{m\zeta\gg 1}\sim \zeta^{3/4}\exp[\pm i\frac{\pi m (2\xi)^{1/2}}{\zeta^{1/2}}] \mbox{,}
\end{equation}
thus, from Eq. (30) we finally obtain
\begin{equation}
\psi_{m\zeta\gg 1}\sim \frac{\phi(\zeta)}{\zeta}\sim \zeta^{-1/4}\cos(\frac{\pi m(2\xi)^{1/2}}{\zeta^{1/2}}+\pi/4) \rightarrow 0 \mbox{,}
\end{equation}
moreover,
\begin{equation}
\psi_{m\zeta\sim 1}\sim(m\zeta-1)^{1/2}\phi(\zeta)\sim O(1) \mbox{,}
\end{equation}
which are the required results in order to apply the quasiclassical formulae
to the solutions of the Wheeler-DeWitt equation.
The barrier (region $R_{I}$) stands for
$1/m \leq \zeta<2/\xi$
while the asymptotic region (namely, $R_{II}$)
is the interval $2/\xi< \zeta < \infty$.

Moreover, WKB solutions of Eq. (23)  may be expressed as
\begin{equation}
\psi(a)=\frac{C a^{p/2}}{V(a)^{1/4}}\exp(- \int V(a)^{1/2} da ) \mbox{,}
\end{equation}
or, for $p=1$,
\begin{equation}
\psi_{I}(a)\approx \frac{2^{1/2}}{\cos^{-1}[\xi/2(m-a)]^{1/2}}
\exp(-\int V(a)^{1/2} da) \mbox{,}
\end{equation}
\begin{equation}
\psi_{II}(a)\approx \frac{2^{1/2}}{\cosh^{-1}[\xi/2(a-m)]^{1/2}}
\cos(\int (-V(a))^{1/2} da+\frac{\pi}{4}) \mbox{,}
\end{equation}
they  have the expected behaviour for the values $am\sim 1$ and  $am\ll 1$.
The constant, $C\approx 2 $, is given by the approximated
normalisation of the wave function between the limits $a=0$ and $a=m$,
we also used $m\gg 1$.
In our real tunneling configuration we only have a  free asymptotic region,
$R_{II}$.
We obtain the WKB barrier penetration rate  simply by
\begin{equation}
\Gamma(m,\xi)=\exp\{-2 \int_{0}^{m-\xi/2} V(a)^{1/2} da\} \mbox{,}
\end{equation}
in the limit $\xi\rightarrow 0$ we get
\begin{equation}
\Gamma(m,0)=\exp\{-2\lim_{\xi\rightarrow 0}\int_{0}^{m-\xi/2} V(a)^{1/2} da\}=
\exp(-\pi m^{2})=\exp(-\pi q^{2}) \mbox{.}
\end{equation}

On the other hand, we only expect magnetic charged wormholes to result from this tunneling
configuration (the electric ones  should decay, also spontaneously,
due to the presence of virtual lower charged particles in vacuum).

More concerning is the fact that
the presence of wormholes causes the structure of spacetime to fail
since they represent geodesically complete spacetimes. This
dangerous feature may originate doubs about the nature of the present
calculation; on the other hand, we can also estimate how strong
this violation of causality is upon suming the rates of production
of all possible magnetic charges, i.e., recalling Dirac's relation
$q_{n}=n/2\sqrt{\alpha}$, $\alpha$ standing for the value of the
electrodynamical fine structure constant,
\begin{equation}
\Gamma(\alpha)=\sum_{n=1}^{\infty}\exp(-\frac{\pi n^2}{4\alpha}) \mbox{;}
\end{equation}
this series is known in the theory of the Riemann's Zeta function
\cite{kn:Brudern}.
It can be expressed as
\begin{equation}
\Gamma(\alpha)=\sqrt{\alpha}-\frac{1}{2}+2\sqrt{\alpha}\Gamma(\frac{1}{16\alpha})\mbox{;}
\end{equation}
thus, it behavies as the squared root of the fine structure constant
when  $\alpha\sim O(1)$ but it goes fairly to zero for small values of its
argument, i.e.,
\begin{equation}
\Gamma_{\alpha\ll 1}\approx \exp(-\frac{\pi}{4\alpha})\ll 1 \mbox{,}
\end{equation}
but
\begin{equation}
\Gamma_{\alpha\sim 1}\approx \sqrt{\alpha}-\frac{1}{2}\sim O(1) \mbox{.}
\end{equation}
Therefore, given the  actual value of $\alpha$, causal violations,
though they might  exist, will cause no real damage to the
underlying spacetime description.
In spite of this, wormholes could have been originated in a bigger
proportion in the very early phases of the Universe (when we
expect a sizable value of $\alpha$); this might well be of
foundamental importance when considering other  quantum physical
properties during this phase.

\section{Summary and conclusions}

We have obtained a simpler interpretation of
the quantum theory of Lorentzian wormholes
connecting two Reissner-Nordstr\"om spacetimes.
This is possible since
the semiclassical approximations to the exact theory are in
direct correspondence
with   the solutions of a more simpler theory whose
Wheeler-DeWitt equation exhibits a potential barrier;
an interpretation  of the solutions  can,  therefore,
be given in terms of
real tunneling configuration of spacetimes
(in a way enterely equivalent
to the creation "ex nihilo" of the Universe in the cosmological case).
Here, however, the semiclassical barrier is very close to a
singularity in the potential. In spite of that, we have shown that the pole
can be mapped to infinity within an asymptotically
free region of the configuration space. The latter allows for the
evaluation of the tunneling rate for a wormhole to be created from nothing
in the extreme  Reissner-Nordstr\"om case. Extreme Reissner-Nordtr\"om
spacetime seems to be the favoured selection since it also allows for the
complex conjugation invariance of the exact Visser's theory.
It might also
be related with the existence of strong limitations for a four dimensional
(Euclidean) topology to change\cite{kn:Sorkin} \cite{kn:Gibbons-K}.

The fact that the topology could change in this way has, however, two
necessary  requirements; the first one is that
it must be regulated by a negative (even small) mass parameter. This agrees
with Tipler's theorem (the example that we have put forward  represents,
therefore, a very pedagogical model of the possible limitations of the
classical theory of gravity). The second  requirement is that the
violations of causality arising from the existence of wormholes
should not damage  the underlying predictive spacetime description.
However, we have also shown
that the rate of creation of these structures
is actually negligible, in the magnetic case,
given the  value of the electrodynamical fine structure constant (but
that it might not be  so small in some earlier phase of the Universe).

\section{Acknowledgements}

The author wants to thank the Department of Physics of the University of
Freiburg for his hospitality. This work is supported by a postdoctoral
grant from the Spanish Ministry of Education and Culture and the research
project C.I.C. y T., PB 94-0194.

\end{document}